\documentstyle[aaspptwo,psfig]{article}

\def\etal{et al. }

\begin{document}

\title
{\bf On The Relationship Between Age and Dynamics in Elliptical Galaxies} 

\author{Duncan A. Forbes, Trevor J. Ponman}
\affil{School of Physics and Astronomy, University of Birmingham,
Birmingham, B15 2TT}
\affil{Email: forbes, tjp@star.sr.bham.ac.uk}

\begin{abstract}

Galaxy age estimates 
(mostly from spectroscopy of the central regions) are
now available for many early type galaxies. In a previous paper 
[Forbes, Ponman \& Brown 1998, ApJ, 508, L43], we showed that the 
offset of galaxies from the fundamental plane depends on galaxy age.
 
Here, using the same sample of 88 
galaxies, we 
examine the scatter about the Faber--Jackson (F--J) relation,
and find that 
a galaxy's position relative to this relation 
depends on its age. In particular, 
younger ellipticals are systematically brighter in M$_B$ and/or have a
lower central velocity dispersion ($\sigma _o$).  
The mean relation corresponds to galaxies that
are $\sim$ 10 Gyr old. 

We attempt to reproduce the observed trend of the F--J residuals
with age using two simple models. The first assumes that galaxy age is
tracing the last major star formation event in an elliptical galaxy. 
We assume that this starburst was instantaneous, centrally located and
involved 10\% of the galaxy by mass. 
The fading of this burst changes 
the M$_B$ component of the F--J residuals, with time.  
Such a model was very successful at
reproducing the B--V and Mg$_2$ evolution reported in our previous paper,
but is unable to reproduce the strength of the F--J trend.  
A second model 
is required to describe age--correlated changes in galaxy dynamics. 
Following expectations from cosmological simulations, we assume 
that $\sigma _o$, for a galaxy of a given mass, 
scales with the epoch of galaxy formation, i.e. with the 
mean density of the Universe. Hence recently formed ellipticals have
systematically lower velocity dispersions than old ellipticals.  
We find that a combination of these two models provides a good match to the
change in F--J residuals with galaxy age. 
This suggests that young ellipticals will have subtly different 
dynamical properties to old ellipticals. 

We also find 
that there is {\it not} a strong relationship between a
galaxy's age and its luminosity for our sample.  
This suggests that the {\it tilt} of the
fundamental plane is not totally driven by age.

\end{abstract}


\section{Introduction}

Hierarchical Clustering and Merging (HCM) 
is the generally accepted paradigm governing the formation and evolution of
elliptical galaxies. In this paradign, elliptical galaxies are formed by the
major merger of disk systems and their associated dark matter halos. This
scenario for forming ellipticals was suggested in the 
seminal work of Toomre \& Toomre (1972), and is often referred to as the 
`merger hypothesis'.  

This hypothesis has gained much support over the years (see 
review by Barnes \& Hernquist 1992). Simulations can now be used to
track the stellar, gaseous and dark matter evolution of a merger from
initial contact, through coalescence of the two nuclei, to the resulting
remnant -- an elliptical galaxy. On the observational side, this simple
evolutionary sequence is more difficult to follow. The main difficulty is
one of correctly identifying galaxies at different stages of the
evolutionary sequence. 

For the first 1--3 Gyrs after a merger the tidal tails produce a clear
signature of a disk--disk merger, and acts as a `clock' for age--dating
the system. Examples of post--tail merger remnants are
harder to identify. Indeed the lack of clear candidate protoellipticals has 
been called the `King gap' after I. King, who pointed out this
potential missing link in 1977. 
Balmer lines, associated with
the merger--induced starburst, offer another clock. 
However after $\sim$ 3 Gyr they have
weakened to the point that metallicity effects are comparable to age in
dictating the strength of the line. 
This is the so--called age--metallicity degeneracy. This degeneracy has meant
that ellipticals could not be uniquely age--dated from stellar
spectra. However, recently 
new spectroscopic observations and models have broken this degeneracy 
(e.g. Worthey 1994; Trager \etal 1999). The number of ellipticals
with high quality nuclear spectral indices on the Lick/IDS system has grown
rapidly in the last year. Combined with evolutionary model tracks, these
indices can be used to estimate the age of the last major starburst in 
elliptical galaxies. 

In an earlier paper (Forbes \etal 1998, Paper I) 
we used such age estimates for 88 early type galaxies
to investigate the scatter about the fundamental plane. As recently as
1996, J$\o$rgensen \etal (1996) found no 
dependence of the FP residuals on galaxy ellipticity, isophotal shape or 
relative disk contribution. 
They concluded that ``It
remains unknown what the source of [the FP] scatter is''. 
We showed that the FP residuals correlate strongly with galaxy age. 
In addition, the deviation of a galaxy from the 
mean scaling relations of B--V colour and Mg$_2$ line index with mass,
also correlates with age. The data are consistent with the steady reddening,
increasing line strength and global fading of an ageing burst of star
formation. This reinforces the view that galaxy age estimates are tracing
the last major burst of star formation, which has been induced by a gaseous
merger event. 

Here we focus on dynamical effects in evolving elliptical galaxies. In
particular we examine the location of ellipticals with age estimates with
respect to the Faber--Jackson scaling relation.

\section{A Sample of Galaxies with Age Estimates}

In Table 1 we give the basic properties of the 
88 early type galaxies which have age estimates available from the
literature. We have included all galaxies 
from the study of Kuntschner \& Davies (1998) 
in the Table, but we have excluded NGC 1399 from our analysis 
as it is a cD galaxy. From the large compilation of 
Prugniel \& Simien (1996), we list the galaxy name, distance
(which includes Virgocentric and Great--Attractor correction terms
for H$_{\circ}$ = 75 km s$^{-1}$ Mpc$^{-1}$), total B band magnitude, 
central velocity dispersion and residual from the fundamental plane, i.e. 
R($\sigma_o, M_B, \mu_e) = 2log(\sigma_o) + 0.286M_B + 0.2\mu_e - 3.101$. 
For the two galaxies which are not part of the Prugniel \& Simien list 
(i.e. the elliptical NGC 6127 and the
S0 NGC 507), we used Faber \etal (1989) for their properties. 
Galaxies in our sample have a 
typical M$_B$ $\sim$ --20.5 and distance of $\sim$30 Mpc, with 3/4 in low
density environments and 1/4 located in clusters. 

If available, age estimates come from recent 
spectroscopic estimates from Trager \etal (1999), Kuntschner \& Davies
(1998), Tantalo \etal (1998) and Mehlert \etal (1997). Where these were not
available we used the youngest `morphological' age estimate from Schweizer \&
Seitzer (1992). The reference for each age estimate is also given in
Table 1. 

These age estimates require some comment. Firstly, the spectroscopic
estimates come from the central regions and are luminosity weighted, so
a centrally located starburst will dominate the spectral lines and hence the
age estimate. Trager (1997) has shown that even a 10\% by mass young
starburst occurring at the galaxy centre dominates over light of the
old stellar population. Thus the age estimates, in the ideal
situation of two distinct stellar populations, will largely 
reflect that of the young
starburst. Secondly, the morphological ages have been `calibrated' to a
starburst model by Schweizer \&
Seitzer (1992). In this sense, they should also reflect the time since the
last major starburst. If we separate our sample into spectroscopic and
morphological ages, both subsamples reveal the trend found in paper I,
i.e. FP residuals correlate with galaxy age. 

In order to include some younger systems, we have included 
data for four good candidates for $\sim$ 1 Gyr remnants of a major gaseous
merger. In Table 2 we list their galaxy names, distances, absolute B
magnitudes, velocity dispersions, FP
residuals, age estimates and reference sources. 

\section{Galaxy Age and Luminosity}

In Paper I we showed that the {\it scatter} about the FP was correlated
with galaxy age. It is also important to determine whether the {\it tilt}
of the FP is also strongly affected by age. If the FP tilt 
results in part from age effects 
then we might expect to see a relationship between the age of a 
galaxy and its luminosity.

The initial Lick sample of about 40 galaxies, suggested that low
luminosity ellipticals were systematically younger than their high
luminosity counterparts (Faber \etal 1995). This is opposite to the trend 
expected from HCM models (Kauffmann \& Charlot 1998; Baugh \etal 1998). 
In Fig. 1 we show the absolute B band magnitude versus galaxy age for our
sample. 
We also show the mean age and the standard deviation of the data 
within bins of 1 magnitude. This 
reveals a weak trend for increasing mean age with increasing luminosity,
but given the large scatter our sample is consistent with no age trend. 
One interesting feature is that only giant 
ellipticals are very old.   

Can this weak trend explain the tilt of the FP ? Prugniel \& Simien (1996)
found that the variation of 
mass-to-light ratio along the FP needed to explain the tilt required   
M/L$_B$ $\propto$ L$_B^{0.27}$. For our data set, 
the factor of 40 change in L$_B$ (i.e. 4 mags) requires 
a change in M/L$_B$ by a factor of 2.7 to fully account for the tilt. 
Using the Bruzual \& Charlot (1993)
models we can estimate the change in global M/L$_B$ for a 10\% by mass
starburst embedded in an old population (i.e. as used in Paper I). 
From a mean starburst age of 5.8 Gyrs (for the low luminosity galaxies) to 7.6
Gyrs (for the high luminosity galaxies), the M/L$_B$ ratio changes by a
factor of 1.03. Thus age variations may be responsible for a trend in 
M/L$_B$ along the FP of this factor, which 
is much less than the factor of 2.7 change 
needed to fully explain the FP tilt. We conclude that 
the tilt of the FP, unlike the scatter about the FP, is not driven by 
galaxy age.

\section{The Evolution of Merger--Remnants}

Galaxy scaling relations such as B--V or Mg$_2$ vs M$_B$ or 
log~$\sigma_o$ are largely sequences of metallicity vs galaxy mass 
(e.g. Kodama \etal 1998). 
In Paper I we showed that the scatter about these relations 
correlate with galaxy age. In particular, after a merger--induced starburst 
the system fades, reddens and has
an increasing Mg$_2$ line index that is consistent with an ageing burst of
star formation. 
This simple model provided a reasonable match to the age--related trends in
these scaling relation residuals without the need for any  
change in the
central velocity dispersion (i.e. dynamical evolution) following 
the merger event. 
Ideally we would like to predict the
evolutionary changes within the full fundamental plane, however the 
difficulty lies in predicting the change of $\mu_e$ (or $r_e$) with time. 
This would require specific assumptions regarding 
the radial surface brightess profile of the old stellar population and 
how the central starburst size changes, and fades, with respect to it. 
Future simulations may indeed be able to address this issue. In this paper,
we examine the overall luminosity evolution of a fading starburst combined
with dynamical effects. 

The Faber--Jackson relation (Faber \& Jackson 1976) is another 2D scaling
relation that relates absolute magnitude (or luminosity) to central
velocity dispersion. It is close to an edge--on view of the FP. 
It was also one of the first scaling relations discovered
for elliptical galaxies and was used by Lake \& Dressler (1986) to study
the dynamical properties of merger remnants. 
The best fit to 229 early type
galaxies with M$_B$ $<$ --18 from the compilation of Prugniel \& Simien
(1996) gives\\ 
log $\sigma _o = 0.102 (-M_B) + 0.243$. 

\noindent
In Fig. 2 we show the residuals about this fit, i.e.
R$(\sigma _o,M_B)={\rm log} \sigma _o - 0.102 (-M_B) - 0.243 $, 
against galaxy
age for our sample of 88 early type galaxies and the 4 candidate
merger--remnants. The mean residual value for our sample is
R$(\sigma _o,M_B)$ = --0.01 indicating that, on average, it is well
represented by the relation derived from the larger 
Prugniel \& Simien sample. 
Overlaid on the data are two model evolutionary tracks and the combination
of the two.  
The short dashed line is the track for the model used in
Paper I, i.e. a 10\% (by mass) central starburst, of solar metallicity, 
superposed on
an non--evolving old stellar population. 
Such a model, of a fading central
starburst, provided a good representation of the residuals for the four 
scaling relations given in
Paper I. This is not the case for the Faber--Jackson residuals, where 
young ellipticals generally 
lie below the model line and old ellipticals lie above
it. A stronger burst, e.g. one involving 50\% of the galaxy's mass, would
provide a steeper curve, which would give a better fit to the data.
However, such a strong burst would give much stronger 
evolution in colour and line strength 
than is observed (see Paper I), as well as being improbably large 
when compared to observed starbursts.
The burst model in Paper I assumed solar
metallicity. Burst models with non--solar metallicities have only 
a minor impact on the Faber--Jackson track, and therefore
also fail to reproduce the trend in the data. 

If post--merger fading cannot explain the magnitude of
the trend in the Faber--Jackson residuals with age, then this suggests that we
should examine the role of dynamics.
All burst models used thus far, have assumed no change in central velocity 
dispersion with age, for a galaxy of given mass. 
We now relax this assumption, and explore the consequences.

\subsection{Dynamical Trends}

We have seen above that fading of a central starburst, which can explain
the evolution of colour and Mg$_2$ index rather well (see Paper I), 
is able to account
for only about one third of the $\Delta$R$(\sigma _o, M_B)$ 
$\approx 0.3$ dex trend in
Faber--Jackson residuals for galaxies older than 1 Gyr, as 
seen in Fig.2. To explain the bulk of
this trend, on the basis of changes in $\sigma_o$, would require a
systematic change in $\sigma_o$ of approximately 60\% across the plot.
This could be achieved in one of two ways: either (a) the central
velocity dispersion increases with time after a merger in an individual
elliptical galaxy, or (b) old ellipticals (i.e. those formed at high
redshift) formed with systematically higher velocity dispersions (for a
given mass) that their younger counterparts.

Substantial intrinsic dynamical evolution in post--merger ellipticals,
over the long timescales required appears extremely unlikely. Simulations
of galaxy merging (e.g. Navarro 1990) indicate that the velocity dispersion
within the effective radius 
relaxes on a timescale $<1$~Gyr after nuclear merger. Infall
of tidal tail material over the next few Gyrs (Hibbard \& Mihos 1995)
cannot extend this timescale
greatly, and in any case will have little impact on the central velocity
dispersion. 

Environmental effects could certainly operate over a longer timescale.
Most ellipticals reside in groups and clusters, and may therefore be
subject to a variety of external influences.
Accretion of satellite galaxies might cause long term
changes in $\sigma_o$ without resetting the post--merger age, but this
should act in the direction of {\it reducing} $\sigma_o$, relative to the
mean Faber--Jackson relation (Hernquist, Spergel \& Heyl 1993).

Another possibility is that $\sigma_o$ does not evolve, but that
ellipticals formed earlier have higher $\sigma_o$ values. The HCM 
models predict that structures collapsing at earlier epochs will
have a higher mean density (e.g. Navarro, Frenk \& White 1997), and hence from
the virial theorem, will have higher velocity dispersion for a given mass.
A number of numerical and theoretical studies (e.g. 
Cole \& Lacey 1996; 
Navarro, Frenk \& White
1997;  Tormen, Bouchet \& White 1997; Salvador-Sole,
Solanes \& Manrique 1998) suggest that it is reasonable to regard the
evolution of dark matter--dominated halos as proceeding via a combination
of merging and accretion. The mean density of a halo, which is set at the
time of a substantial merger event, is a multiple of the mean density of
the Universe at that epoch. Subsequent accretion of material onto the
halo does not disrupt the density distribution in its inner regions, so
that the central velocity dispersion will remain essentially constant, until it
is reset by a further major merger event.

Since the dark halos formed in numerical simulations have a universal
form (Navarro, Frenk \& White 1997), which seems to accord well
with observations, the velocity dispersion will, by the virial theorem,
scale in a simple way with system mass and mean density:
$$ \sigma^2 \propto M/R \propto M^{2/3} {\rho}^{1/3} 
\propto M^{2/3} (1+z_f), $$
where M and R are the mass and virial radius of the system, and
${\rho}$ is its mean density, which scales (Lacey \& Cole 1993) with
the critical density of the Universe at the epoch ($z_f$) of its last
major merger event. 
Hence a galaxy of a given luminosity (and therefore mass) will have a
velocity dispersion which depends upon its `formation' epoch as $\sigma
\propto (1+z_f)^{1/2}$. 
Assuming that the 10\% starburst coincides with the final merger which
assembled the galaxy, the 
measured post--merger ages 
provide us with the value of $z_f$ required to evaluate
the strength of the velocity dispersion trend predicted by this
equation. The resulting curve relating age to velocity dispersion is
cosmology-dependent, via the relationship between look--back 
time and redshift.

As an example of the magnitude of the effects which are obtained, we have
overlaid in Fig. 2 (long dashed line) the variation in Faber--Jackson residuals
which result from a flat $\Lambda$CDM Universe with $\Omega_m=0.3$ and
$\Omega_\Lambda=0.7$ for H$_{\circ}$ = 60 km s$^{-1}$ Mpc$^{-1}$. The
model has been normalised to zero residual at 10 Gyr, as Paper I showed
that ellipticals lie on the fundamental plane if they have an age of
$\sim$ 10 Gyr. 
Changing the
Hubble constant has a substantial effect on the model track, e.g. 
for H$_o$ = 75 the track 
lies about 0.1 dex lower around 1 Gyr, and is slightly steeper than the
H$_o$ = 60 track for ages older than 10 Gyr. 
The dynamical trends match the overall magnitude of
the residuals fairly well. But from paper I, we know that there is also an
evolutionary effect on M$_B$ from the fading starburst. This is shown
in Fig. 2 by the short dashed line. The combination of the two effects 
(i.e. dynamical trends plus a burst model) are
given by a solid line, which does a reasonable job of describing the trend
of F--J residuals with age. 

In order for this dynamical effect to be acceptable as an explanation
for the bulk of the Faber--Jackson residuals, it is essential that
it should not conflict with the success of the burst model
in explaining the correlations involving colour
and Mg$_2$ index, which were discussed in Paper I. In practice, the 
coefficients of log~$\sigma _o$ in the relations with B--V and Mg$_2$
explored in the Paper I are small ($\sim0.15$), so that 60\% changes
in log~$\sigma _o$ result in only minor perturbations
to the successful model fits obtained with the fading burst model
under the assumption of no dynamical trends.

\section{Concluding Remarks}

Our main result concerns the scatter about the Faber--Jackson (F--J) 
relation for
elliptical galaxies. We found that 
the scatter about this scaling relation is correlated with the post--merger
age of the galaxy. Young 
ellipticals are systematically brighter in the B band and/or have a
lower central velocity dispersion than the mean relation.  
The mean relation corresponds to galaxies that
are $\sim$ 10 Gyr old. 

We have also shown that we can reproduce the observed trend of the 
F--J residuals with age using two simple concepts. 
First, we assume that the galaxy has undergone an 
instantaneous, centrally located starburst 
involving 10\% (by mass) of the galaxy.  
Such a model was very successful at
reproducing the B--V and Mg$_2$ evolution reported in our earlier paper, 
but we find that it can not adequately explain the F--J residuals with
time. We propose that the bulk of the trend in F--J residuals results from
a systematic variation in the internal dynamics. 
In particular, the central velocity dispersion of a newly formed 
elliptical scales with the mean density of the Universe. 
Such a change in mean density moves a galaxy around on the fundamental
plane, but in the F--J projection of the plane it moves a galaxy away from
the mean relation. The F--J offsets therefore contain information about the
epoch of elliptical galaxy assembly, which may be some time after the bulk
of its stars were actually formed. 
The strength of the dynamical trend observed favours a low value of the
Hubble constant (i.e.  $\le$ 75 km
s$^{-1}$ Mpc$^{-1}$). 
We find that a combination of this effect and starburst fading 
provides a good match to the
change in F--J residuals with galaxy age. The success of this model
suggests that the derived starburst ages measure the time since the
galaxy we see was assembled in essentially its present form. This in turn 
confirms the important role of recent dissipational 
mergers in the formation of elliptical galaxies. 

Our results lend support to the view that the internal 
dynamical properties of an elliptical galaxy depend on when the galaxy was
formed. 
The trend in F--J residuals we observe could have been predicted on the
basis of HCM models, but as far as we are aware, they have not been
explicitly mentioned in the literature. 
It follows that, 
if the merger of two spirals indeed produces an elliptical
galaxy, then today's ongoing mergers will form an elliptical 
galaxy subtly different from old ellipticals found in nearby clusters. 
We might also expect studies of the fundamental plane at redshifts z $\sim$
1 to begin to reveal dynamical differences to local ellipticals. 

\newpage

\noindent{\bf Acknowledgments}\\
We thank R. Brown, P. James and A. Terlevich for help and useful 
discussions. We also thank S. Trager for allowing us to use his results 
prior to publication, and the referee (C. Lacey) 
for several suggestions which have
improved the paper.\\

\noindent{\bf References}\\
Barnes, J. E., \& Hernquist, L. 1992, ARA\&A, 30, 705\\
Baugh, C. M., Cole, S., Frenk, C. S., Lacey, C. G. 1998, ApJ, 498, 504\\
Bica, E., \& Allion, D. 1987, A\&AS, 70, 281\\
%Bender, R. 1990, in The Dynamics and Interactions of Galaxies, ed. R. Wielen 
%(Berlin:Springer), 232\\
%Bender, R., Burstein, D., \& Faber, S. M. 1992, ApJ, 399, 462\\
%Bender, R., Burstein, D., \& Faber, S. M. 1993, ApJ, 411, 153\\
%Bruzual, G. A., \& Charlot, S. 1993, ApJ, 405, 538\\
%Carollo, C. M., Franx, M., Illingworth, G. D., \& Forbes, D. A. 1997, 
%ApJ, 481, 710\\
Cole, S., \& Lacey, C. 1996, MNRAS, 281, 716\\
%de Carvalho, R. R., \& Djorgovski, S. 1992, ApJ, 389, L49\\
%Djorgovski, S., \& Davis, M. 1987, ApJ, 313, 59\\
%Dressler, A., Lynden-Bell, D., Burstein, D., Davies, R. J., Faber, S. M., 
%Terlevich, R. J., \& Wegner, G. 1987, ApJ, 313, 42\\
Faber, S. M., \& Jackson, R. E. 1976, ApJ, 204, 668\\
Faber, S. M., \etal 1989, ApJS, 69, 763\\ 
Faber, S. M., Trager, S., Gonzalez, J., Worthey, G. 1995, in Stellar
Populations, eds P. C. van der Kruit and G. Gilmore (Dordrecht: Kluwer), 249\\
%Faber, S. M., Dressler, A., Davies, R. L., Burstein, D., Lynden-Bell, D., 
%Terlevich, R. J., \& Wegner, G. 1987, in Nearly Normal Galaxies, ed. S. M. 
%Faber (New York:Springer-Verlag), 175\\
%Forbes, D. A., Franx, M., \& Illingworth, G. D. 1995, AJ, 109, 1988\\
Forbes, D. A., Ponman, T. J., \& Brown, R. J. N. 1998, ApJ, 508, L43 (Paper
I)\\
%Gonzalez, J. J. 1993, Ph.D Thesis, University of California, Santa Cruz\\
%Gregg, M. D. 1992, ApJ, 384, 43\\
%Guzman, R., \& Lucey, J. R. 1993, MNRAS, 263, L47\\
%Guzman, R., Lucey, J. R., \& Bower, R. G. 1993, MNRAS, 265, 731\\
%Hibbard, J. E., \& van Gorkom, J. 1996, AJ, 655, 111\\
Hernquist, L., Spergel, D. N., \& Heyl, J. S. 1993, ApJ, 416, 415\\
Hibbard, J. E., \& Mihos, J. C. 1995, AJ, 110, 140\\ 
J$\o$rgensen, I., Franx, M., \& Kjaergaard, P. 1996, MNRAS, 280, 167\\
Kauffmann, G., \& Charlot, S. 1998, MNRAS, 294, 705\\ 
%Kauffmann, G., \& Charlot, S. 1998b, astro-ph/9810031\\
Kodama, T., Arimoto, N., Barger, A. J., Aragon-Salamanca, A. 1998, A\&A,
334, 99\\
Kuntschner, H., \& Davies, R. L. 1998, MNRAS, 295, 29\\
Lacey, C. \& Cole, S. 1993, MNRAS, 262, 627\\
Lake, G., \& Dressler, A. 1986, ApJ, 310, 605\\
Mehlert, D., Bender, R., Saglia, R. P., \& Wegner, G. 1997, astro-ph/9709295\\
%Mihos, J. C., \& Hernquist, L. 1994, ApJ, 431, L9\\
Navarro, J. F. 1990, MNRAS, 242, 311\\
Navarro, J. F., Frenk, C. S., \& White, S. D. M. 1997, ApJ, 490, 493\\
Prugniel, P., \& Simien, F. 1996, A\&A, 309, 749\\
%Renzini, A., \& Ciotti, L. 1993, ApJ, 416, L49\\
%Schweizer, F., Seitzer, P., Faber, S. M., Burstein, D., Dalle Ore, C. M., \& 
%Gonzalez J. J. 1990, ApJ, 364, L33\\
Salvador-Sole, E., Solanes, J. M., \& Manrique, A. 1998, ApJ, 499, 542\\
Schweizer, F., \& Seitzer, P. 1992, 104, 1039\\
Tantalo, R., Chiosi, C., \& Bressan, A. 1998, A\&A, 333, 419\\
Toomre, A., \& Toomre, J. 1972, ApJ, 178, 623\\
Tormen, G., Bouchet, F. R., \& White, S. D. M. 1997, MNRAS, 286, 865\\
Trager, S. C. 1997, Ph.D Thesis, University of California, Santa Cruz\\
Trager, S. C., Faber, S. M., Gonz\'alez, J. J., \& Worthey, G. 1999, in
preparation\\ 
Whitmore, B. C., Miller, B. W., Schweizer, F., \& Fall, S. M. 1997, AJ,
114, 1797\\ 
Worthey, G. 1994, ApJS, 95, 107\\
%Zepf, S., \& Silk. J. 1996, ApJ, 466, 114\\

\begin{figure*}[p]
\centerline{\psfig{figure=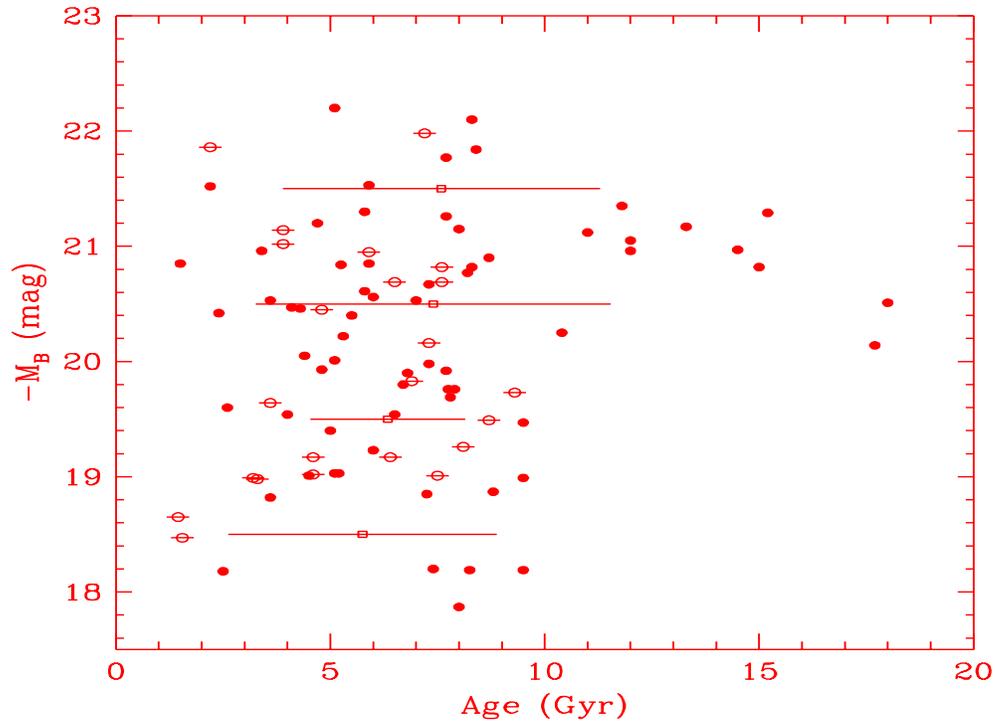,width=400pt,height=400pt}}
\caption{\label{fig1}
Absolute B band magnitude versus galaxy age. Ellipticals are represented by
filled symbols, S0s by open circles with a horizontal line. 
The mean age and standard deviation of the data within 1 mag. 
bins is represented by the
open squares and horizontal lines.  
Our sample shows little, if any, trend of galaxy luminosity with age 
suggesting that the FP tilt is {\it not} strongly dependent on galaxy
age. 
}
\end{figure*}

\begin{figure*}[p]
\centerline{\psfig{figure=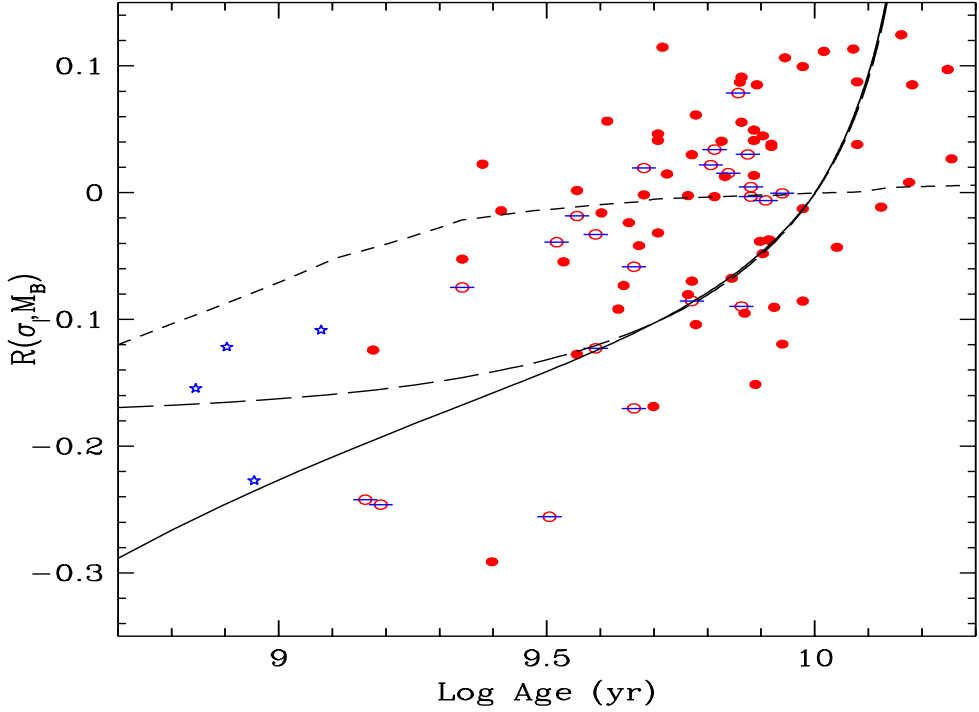,width=400pt,height=400pt}}
\caption{\label{fig2}
Residual from the Faber--Jackson relation versus galaxy age. 
The same symbols are used as in Figure 1, with late stage mergers shown by 
stars. The position of NGC 3610, with R$(\sigma _o, M_B) = -0.66$, is not
shown on the plot. 
The short dashed line shows the evolutionary track for a galaxy 
which has undergone an instantaneous, solar
metallicity starburst involving 10\% of the total galaxy mass with no
change in velocity dispersion. 
The long dashed line shows the
dynamical effects predicted for a  
$\Omega_m=0.3$, $\Omega_\Lambda=0.7$ and H$_o$ = 60 
$\Lambda$CDM cosmology. The solid line shows the combined effects of the
two dashed lines.   
See text for details. 
The evolution from late stage mergers to old ellipticals is 
generally consistent with expected changes in stellar population 
following a merger--induced starburst, including dynamical effects 
due to different conditions at the epoch of formation. 
}
\end{figure*}

\end{document}